# Pulse induced resonance with angular dependent total enhancement of multi-dimensional solid-state NMR correlation spectra

Orr Simon Lusky[1], Amir Goldbourt[1*]

[1]*School of Chemistry, Faculty of Exact sciences, Tel Aviv University, Tel Aviv, Israel.*

[*]amirgo@tauex.tau.ac.il

## Abstract

We demonstrate a new resonance condition that obeys the relation $\Delta\delta = n\nu_R/2$, where $\Delta\delta$ is the chemical shift difference between two homonuclear-coupled spins, $\nu_R$ is the magic-angle spinning speed and $n$ is an integer. This modulation on the rotational resonance recoupling condition is obtained by the application of rotor-synchronous [1]H pulses when at least one proton is dipolar-coupled to one of the homonuclear-coupled spins. We suggest a new experimental scheme entitled 'pulse induced resonance with angular dependent total enhancement' (PIRATE) that can enhance proton-driven spin diffusion by the application of a single [1]H pulse every rotor period. Experimental evidence is demonstrated on the two carbon spins of glycine and on the Y21M mutant of fd bacteriophage virus. Numerical simulations reveal the existence of the resonances and report on the important interactions governing this phenomena.

## Introduction

Solid state NMR (ssNMR) has become a common tool in structural biology studies. Its various merits include the ability to study biomolecules such as proteins, nucleic acids, polysaccharides, and their complexes [1–9]. Moreover, ssNMR methods are able to characterise the dynamics of molecules on a wide range of time scales [10,11]. Such studies rely heavily on efficient magnetisation transfer methods that utilise spin-spin interaction. Experiments such as INADEQUATE [12] and TOBSY [13] utilise scalar couplings to transfer polarisation through bonds between homonuclear spins. However, these interactions are relatively weak, in the order of Hz, and therefore require long mixing times and sufficiently long relaxation times as well. Anisotropic interactions, and particularly the dipolar interaction, are much stronger at short distances (a few ångstroms); however, they are averaged with the sample spinning. At low spinning rates, the anisotropic interactions are only partially averaged, and may be utilised for magnetisation transfer through space. The 'proton driven spin diffusion' (PDSD) [14] experiment is an example of how such transfer can be used, and was successfully demonstrated in many occasions, including the determination of the first protein structure by NMR [15]. Spin diffusion itself can be utilised very efficiently to transfer polarisation between protons in experiments based on the X-H-H-X scheme, where X can be $^{31}$P, $^{13}$C, $^{15}$N, or other nuclei [16,17].

Since PDSD relies on residual dipolar couplings that are not averaged by spinning, its performance deteriorates at elevated spinning frequencies. Further enhancement of magnetisation transfer can be obtained by recoupling; generating an average interaction Hamiltonian that is time independent during the evolution time, cancelling (at least partially) the effect of the MAS. Numerous such pulse sequences have been proposed [18], with most methods relying on the generation of a particular resonance condition that nullifies the MAS time averaging. The resonance conditions can be met by setting the right experimental parameters, as in the simplest case of rotational resonance ($\mathscr{R}$2) [19]. This condition requires fixing the MAS rate according to $\Delta\delta = n\nu_R$, where $\Delta\delta$ is the isotropic chemical shift difference between two homonuclear coupled spins, $\nu_R$ is the MAS rate and $n$ is an integer. When this condition is met, a spinning sideband of one spin overlaps with the isotropic shift of the other spin, and recoupling of the homonuclear dipolar interaction is achieved without the application of any external radio-frequency (rf) pulses.

Rf pulses can also be used for recoupling. Various approaches utilise direct homonuclear polarisation transfer. For example radio frequency driven recoupling (RFDR) [20] utilises rotor-synchronous hard π pulses on the homonuclear-coupled spins to obtain recoupling with a larger bandwidth than $\mathscr{R}$2, while narrowband RFDR has a narrower chemical shift offset dependence that depends on the spacing between the π pulses, and efficiency that similarly to $\mathscr{R}$2 is maximized at $\Delta\delta = n\nu_R$ [21]. SEDRA [22] and nbSEDRA [23] differ from the former only by the fact that recoupling occurs in the transverse plane. Interestingly, finite-pulse RFDR [24] utilise similar π pulses as in RFDR however when the pulses occupy a significant portion of the rotor period (~ 30%) the sequence does not depend on the chemical shift offset any longer and recoupling occurs during the pulses. In the 'homonuclear rotary resonance' (HORROR) experiment the spins are irradiated at an intensity ($\nu_1$) that equals half the rotation speed, $\nu_1 = \nu_R/2$ [25]. This experiment was further improved by sweeping adiabatically through this condition in an experiment termed DREAM, dipolar recoupling enhancement through amplitude modulation [26]. Many homonuclear recoupling experiments can be classified under the symmetry rules of C and R sequences [27], including their supercycled versions [28].

Another approach for achieving homonuclear transfer utilises the coupled proton bath and is based on the aforementioned PDSD technique. In PDSD, the heteronuclear interaction between the protons and the homonuclear coupled X spins (X can be $^{13}$C, $^{15}$N etc.) is only partially averaged by sample spinning, promoting transfer between the X spins. However, active recoupling of the $^1$H-X heteronuclear interaction further enhances polarisation transfer between the X spins. For example, such recoupling was achieved by irradiating on one of the spins with an rf power that equals to an integer of the MAS rate ($v_{1H} = nv_R, n = 1,2$) [29]. This resonance condition is known as 'rotary resonance recoupling', $\mathscr{R}3$, and has been used in the 'dipolar assisted rotational resonance' (DARR) experiment to enhance proton driven spin diffusion [30]. It was further shown that supercycling of the continuous $^1$H recoupling pulse increases the bandwidth and can improve recoupling [31,32]. Further adjustments for spin rates beyond 25-30 kHz have been proposed. In the mixed rotational and rotary resonance' (MIRROR) [33] experiment the rf power applied to the $^1$H pulse is matched to the sum of the spinning speed and the chemical shift offset difference, $v_{1H} = nv_R + \Delta\delta$. Another approach, COmbined R2n(v)-Driven broadband homonuclear recoupling (CORD) [34], utilises a combination of the rotary resonance condition and the HORROR condition thereby increasing the bandwidth of $\mathscr{R}3$-based recoupling. Another mechanism was proposed that enables transfer of magnetisation between homonuclear spins. In the proton-assisted recoupling (PAR) experiment, a single $^1$H spin coupled to two homonuclear spins is sufficient to promote magnetisation transfer if all three spins are irradiated simultaneously at proper conditions [35]. This mechanism was earlier proposed to promote heteronuclear spin transfer in an experiment entitled TSAR, third-spin assisted recoupling [36]. In both cases, Hartman-Hahn ($v_{1H} = v_{1X} \pm v_R$) and $\mathscr{R}3$ ($v_1 = nv_R$) conditions should be avoided, and the experiments are well suited for spinning rates beyond 30 kHz.

Recently we have shown that by applying rotor synchronous π pulses on $^1$H during $^{15}$N-$^{15}$N PDSD transfer, an enhancement is obtained at relatively low spinning speeds (8-10 kHz) [37]. This enhancement allowed us to detect hydrogen-bonds in the full-length RNA extracted directly from MS2 bacteriophage. Particularly, the low gyromagnetic ratio of $^{15}$N, and the dynamic nature of the 3569-bases-long genome required the application of low MAS rates and very long mixing times (up to 16 sec). Addition of the proton pulses was essential for observing inter-nucleotide contacts.

In this paper we explore in detail the effect of $^1$H pulses during proton-driven spin-diffusion. We show that the application of a single $^1$H $\pi$ pulse every rotor period during PDSD enhances polarisation transfer and in addition, generates a new resonance condition that is satisfied when the difference between the chemical shifts of coupled spins equals half integers of the spinning speed ($\Delta\delta = n\nu_R/2$). We term this new condition 'half rotational resonance' ($\mathcal{HR}2$), as it appears to be a generalisation of the $\mathcal{R}2$ condition and fits the modulation frequency of the $^1$H $\pi$ pulses. Using numerical simulations, we demonstrate that the mechanism relies on a single proton that is dipolar coupled to at least one of two homonuclear coupled spins (e.g. $^{13}$C, $^{15}$N). We also show that the usage of different flip angles for the rotor-synchronous pulses creates a trade-off between enhancement of the proton-driven spin-diffusion mechanism and the efficiency at the $\mathcal{HR}2$ condition. Using experiments on fully $^{13}$C labelled glycine as a two-spin model system, and on a $^{13}$C/$^{15}$N uniformly labelled sample of fd-Y21M phage we discuss the potential of this experiment in the design of solid-state NMR experiments at low to moderate MAS rates.

## Materials and Methods

**Sample.** Fully labelled $^{13}$C glycine was purchased from Cambridge Isotope Laboratories, Inc., and was packed into a 4mm $ZrO_2$ MAS rotor. A fully labelled $^{13}$C and $^{15}$N fd-Y21M bacteriophage sample was prepared using our lab protocols [38], and was packed into a 4mm $ZrO_2$ MAS rotor.

**NMR methods.** NMR experiments were carried out on a Bruker Avance III spectrometer operating at 14.1T, equipped with a MAS 4mm probe. A complete list of experimental parameters appears in the supporting information (SI). The chemical shifts of $^{13}$C were externally referenced to Adamantane at 40.48ppm [39]

**Data Analysis.** NMR data were processed using TopSpin3.5. Analysis was performed using TopSpin3.5 and SPARKY version 3.134 [40].

**Numerical Simulations.** The NMR simulation package SIMPSON [41] was used to simulate the application of the pulse schemes on an $X_2H_3$ spin system ($X\equiv^{13}$C in this case, and represents a low-$\gamma$ spin ½). The starting operator was $I_{1z}$-$I_{2z}$, and the detection operator was $I_{2z}$ for all the simulations. Its value was normalised to -1 depicting an ideal inversion and no mixing. The script and simulation parameters appear in the SI.

## Results and Discussion

### Simulations

**Magnetisation transfer with ¹H π pulses:** We chose to simulate an $X_2H_3$ dipolar-coupled spin system, where the X spins are some insensitive nucleus (e.g. $^{13}C$, $^{15}N$). Such a system allows us to explore three different transfer mechanisms, namely proton driven spin diffusion, direct dipolar transfer, and third-spin assisted recoupling. The total Hamiltonian for such a system in the rotating frame (Eq. 1) contains the chemical shifts ($\widehat{\mathcal{H}}_{CSA}$) of spins X and H, the homonuclear ($\widehat{\mathcal{H}}_d^{II}$, $\widehat{\mathcal{H}}_d^{SS}$) and heteronuclear ($\widehat{\mathcal{H}}_d^{IS}$) dipolar interactions with I and S representing X and ¹H spins respectively, and the time dependent radiofrequency Hamiltonian, $\widehat{\mathcal{H}}_{rf}(t)$.

$$(1)\ \widehat{\mathcal{H}}_{X_2H_3}(t) = \widehat{\mathcal{H}}_{CSA}^S(t) + \widehat{\mathcal{H}}_{CSA}^I(t) + \widehat{\mathcal{H}}_d^{SS}(t) + \widehat{\mathcal{H}}_d^{II}(t) + \widehat{\mathcal{H}}_d^{IS}(t) + \widehat{\mathcal{H}}_{rf}(t).$$

The density operator of the two X spins is set to an initial state of $I_{1z}$-$I_{2z}$ and magnetisation transfer is monitored by the detection operator $I_{2z}$, calculated after evolution times that are whole integers of the rotor period. In order to apply the pulse scheme, ideal pulses, and later on finite pulses are applied to the proton channel in the middle of each rotor period. Changing the position of the pulse to the beginning or end of the rotor period had no visible effects on the results.

We initially compared the effect of ¹H 180° pulses by comparing such a scheme to PDSD. Two typical build-up curves are shown in figure 1a, one with the application of the pulses, and one without. Clear enhancement is observed at the chosen conditions, MAS rate of 6 kHz and a chemical shift difference of 5 kHz. We then modified the chemical shift difference between the two X spins and extracted the values of $I_{2z}$ that correspond to maximal magnetisation transfer in the build-up curves. Figures 1b shows the maximal transfer efficiency in absence and presence of 180° pulses. In all recorded spectra the $\mathcal{R}2$ condition was apparent. However, in all cases where 180° pulses were applied, additional maxima were spotted. Those maxima correspond to a new resonance condition obeying the following relation:

$$(2)\ \Delta\delta = \frac{n}{2}\nu_R.$$

It generalises the $\mathscr{R}2$ condition by allowing recoupling at MAS rates that equal the chemical shift difference also by half integers, or 'half-rotational resonance', $\mathscr{HR}2$.

In addition to the enhancement resulting from the $\mathscr{HR}2$ condition, it can be seen that the maximal magnetisation transfer is enhanced for all values of Δδ due to the application of the pulses. This suggests an enhancement of the PDSD effect in addition to the $\mathscr{HR}2$ condition and is therefore applicable to homonuclear correlation experiments. We already demonstrated this effect on $^{15}$N-$^{15}$N correlation spectra of full-length RNA extracted from the MS2 bacteriophage, where the $^1$H pulses enhanced the sensitivity of the experiment allowing us to observe many inter-nucleotide cross-peaks [37].

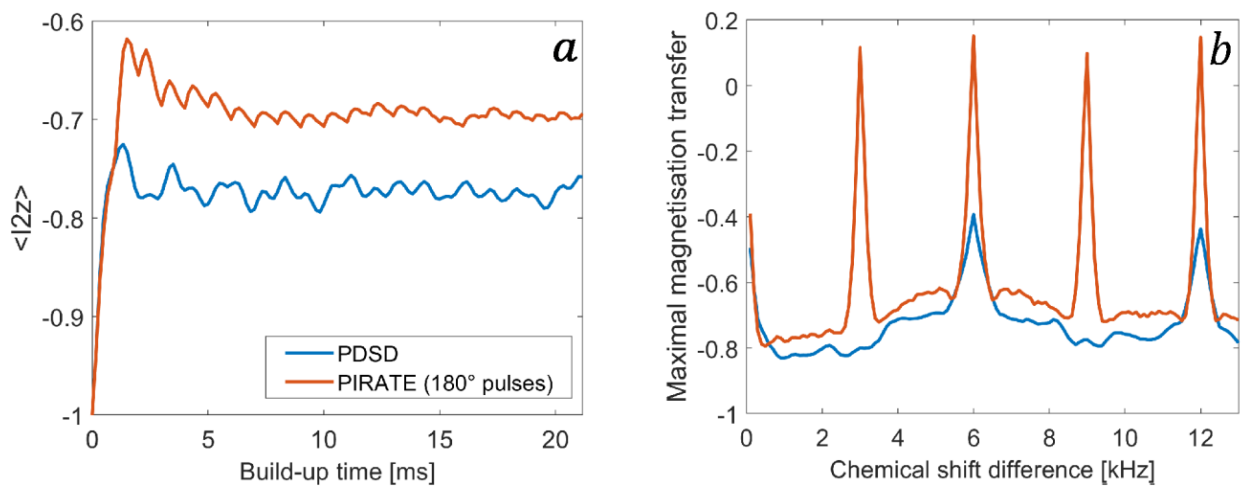

**Figure 1:** (a) Simulations of build-up curves showing the magnetisation of a selectively inverted $^{13}$C spin in a fully dipolar-coupled $^{13}$C$_2$$^1$H$_3$ spin system. The MAS rate was 6 kHz and the chemical shift difference between the two $^{13}$C spins was 5 kHz. The data displayed on the y-axis is normalised so that I$_{2z}$ (0) =-1. (b) Maximal magnetisation transfer as a function of the chemical shift difference Δδ between the two $^{13}$C spins at a MAS rate of 6 kHz. In both plots the blue line represents magnetisation transfer in the absence of pulses (PDSD experiment), whereas the orange line represents the transfer following the application of $^1$H 180° pulses (PIRATE experiment). Here the maximal $^1$H-$^1$H coupling is 55 kHz. In Figure S1 in the SI this is reduced to a typical CH$_2$ group (30 kHz) showing that all resonances have a similar efficiency.

We can identify three different effects in figure 1b; rotational resonance, half-rotational resonance, and proton-driven spin diffusion, all of which exist in the presence of $^1$H 180˚ pulses.

In order to identify the interactions that contribute to the enhancement of PDSD and the appearance of the $\mathscr{HR}2$ resonance, we compared in figure 2 the maximal expectation values of I$_{2z}$ that are obtained by a fully coupled X$_2$H$_3$ spin system (blue curves) to conditions, in which we selectively eliminate or change different parts of the Hamiltonian. The expectation values were extracted from the maximum of the build-up curves, such as those shown figure



S2 of the SI. Initially, the homonuclear proton dipolar interaction term $\hat{\mathcal{H}}_d^{HH}$ was nullified, significantly reducing the PDSD effect (fig. 2a). Yet both $\mathcal{R}2$ and the new $\mathcal{HR}2$ were still present. We can conclude that the $^1$H-$^1$H dipolar couplings are not required for generating the new resonance condition. When $\hat{\mathcal{H}}_d^{HX}$, the heteronuclear dipolar Hamiltonian, was nullified as well (practically removing the protons from the system) recoupling only appeared at $\mathcal{R}2$, with a negligible magnetisation transfer in any other value of Δδ, as seen in figure 2b. In figure 2c $\hat{\mathcal{H}}_d^{HH}$ was reintroduced and $\hat{\mathcal{H}}_d^{HX}$ was limited to the closest X spins only. This results in recoupling that is very similar to a fully coupled system.

In figures 2d and 2e we search for a minimal set of interactions that still retain the $\mathcal{HR}2$ effect. We find that a single $^1$H-$^{13}$C interaction in the presence of $^{13}$C-$^{13}$C coupling is sufficient to introduce the half-rotational resonance transfer condition.

In order to compare this resonance to the PAR sequence, where the homonuclear interaction is not required to promote magnetisation transfer, the $^{13}$C-$^{13}$C homonuclear interaction was nullified, resulting in no magnetisation transfer. This is shown in figure S2i in the SI.

We then modified the interaction strengths. When the $^{13}$C-$^{13}$C coupling was reduced, the contribution of PDSD was reduced for all Δδ but $\mathcal{HR}2$ remained as efficient as before, as seen in figure 2f. The only difference was that the build-up time was longer (compare figures S2a and S2f in the SI). When the orientation of the dipolar couplings between two spin-pairs was matched (indicated in figure 2g as thick coloured lines), the results showed a negligible change in the recoupling profile. Similarly the effect of $\hat{\mathcal{H}}_{CSA}^X$ shown in figure 2h shows a very small decrease in the total magnetisation transfer and only in the presence of very large CSA values (200-500 ppm, atypical for carbons) and no effect on $\mathcal{HR}2$.



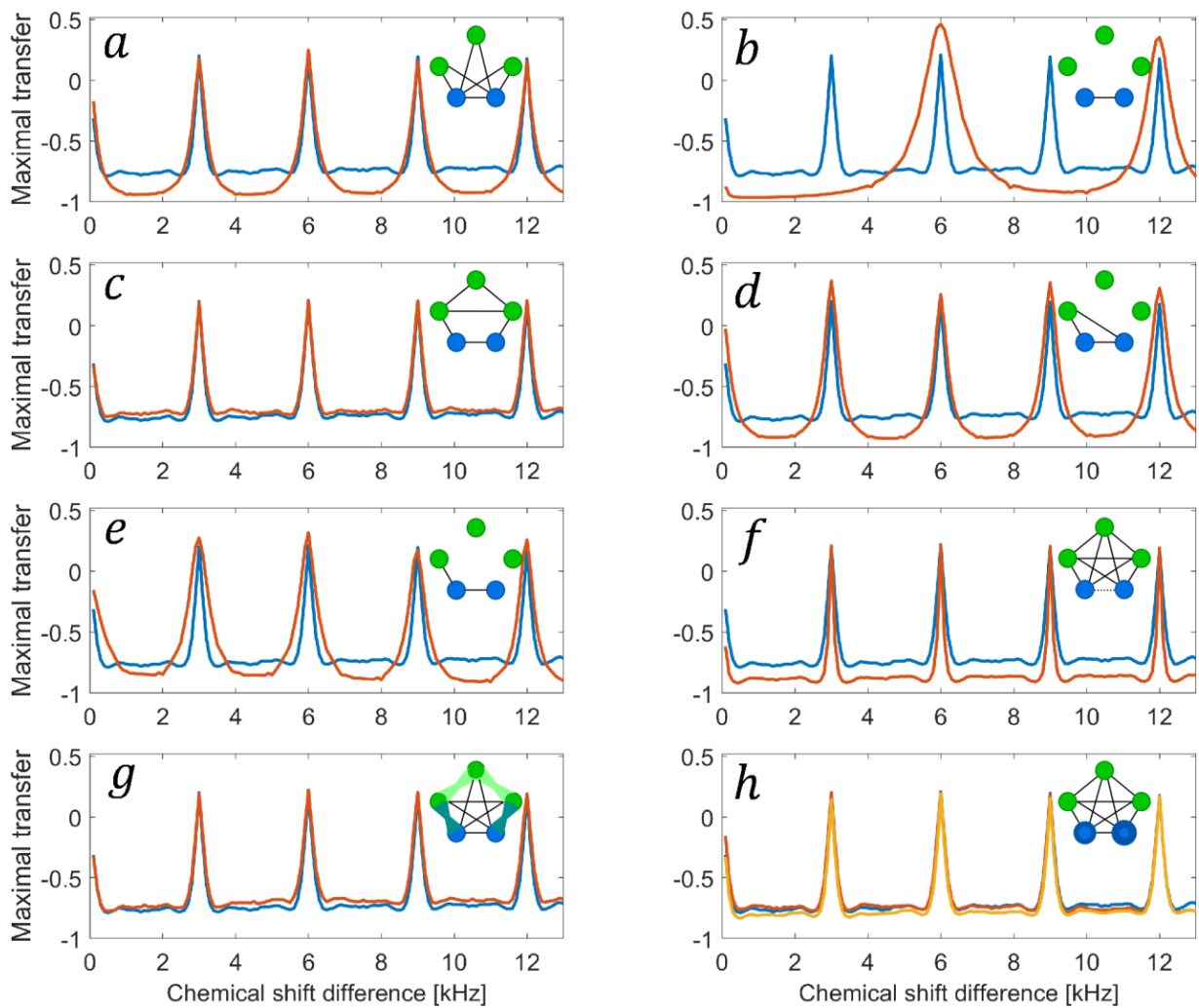

**Figure 2:** Comparisons of the maximal polarisation transfer efficiency from a selectively inverted $^{13}C$ spin to another $^{13}C$ spin in a $^{13}C_2H_3$ spin system ($^{13}C$ in blue, $^1H$ in green) as a function of the chemical shift difference between the $^{13}C$ spins. The MAS rate is 6 kHz in all cases, a single 180° pulse was applied to $^1H$ spins every rotor period, and the coupling values are varied. The maximal transfer was obtained by taking the maximum from build-up curves such as those shown in figure S2 in the SI for particular cases. The blue curve in all graphs represent a fully coupled system with the following dipolar coupling constants: $D_{13C-13C}$ = 2 kHz, $D_{1H-13C}$ = 21.94 kHz, $D_{1H-1H}$ = 25.04 kHz twice and 55.04 kHz once. The orange curve is for (a) $D_{1H-1H}$ = 0 kHz; (b) no protons: $D_{1H-1H}$ = $D_{1H-13C}$ = 0 kHz; (c) Four $D_{1H-13C}$ couplings removed (d) $^{13}C_2{}^1H_1$ spin system with both $D_{1H-13C}$ couplings present; (e) $^{13}C_2H_1$ spin system with a single $D_{1H-13C}$ coupling present; (f) fully coupled system, $D_{13C-13C}$ = 1 kHz; (g) fully-coupled system with orientation dependence ($\beta$ angles of two X-H pairs and two H-H pairs (both $D_{1H-1H}$ = 25.04 kHz) dipolar tensors were set to 60° and 40° respectively) with the same angles of the dipolar interaction tensor; (h) fully coupled system, CSA of $^{13}C$ are 200 ppm/300 ppm in orange, and 200 ppm/500 ppm in yellow. The data displayed on the y-axes are normalised; $I_{2z}(0)$ = -1. No transfer was obtained when $D_{13C-13C}$ = 0 kHz (Figure S2i).

In order to ensure that the magnetisation is transferred solely between $I_{1z}$ and $I_{2z}$, we simulated in three different cases the values of $I_{1z}$, $I_{2z}$, and their sum $I_{1z}+I_{2z}$. As can be seen



in Figure S3 of the SI, the sum remains constantly at zero, thus there is no loss of magnetisation to higher order terms in these cases.

In summary, the simulations suggest that in order to obtain transfer at the half-rotational resonance condition, the minimal set of interactions that is required is defined by a single heteronuclear and a single homonuclear coupling term, that is, an H-X-X system.

**Magnetisation transfer with finite $^1$H pulses:** The dependence of the magnetisation transfer on the duration of the pulse applied on the protons was simulated in Figure 3. A series of different pulse durations were simulated to show the transition from the new condition fulfilment to the HORROR condition. As the pulse length increases, the efficiency of the magnetisation transfer at the new condition decreases. When the pulse occupies 5% of the rotor period the $\mathcal{HR}$2 condition is still apparent although weak, whereas at 10% and higher no enhancement at the resonance condition is apparent. When the length of the pulse equals the rotor period, the pulse becomes a continuous irradiation and the HORROR condition is met. Here also the $\mathcal{HR}$2 condition disappears. The limitation of pulse length can be more easily be met at low MAS rates where the rotor period is longer, or with smaller rotors where high power pulses can be applied.

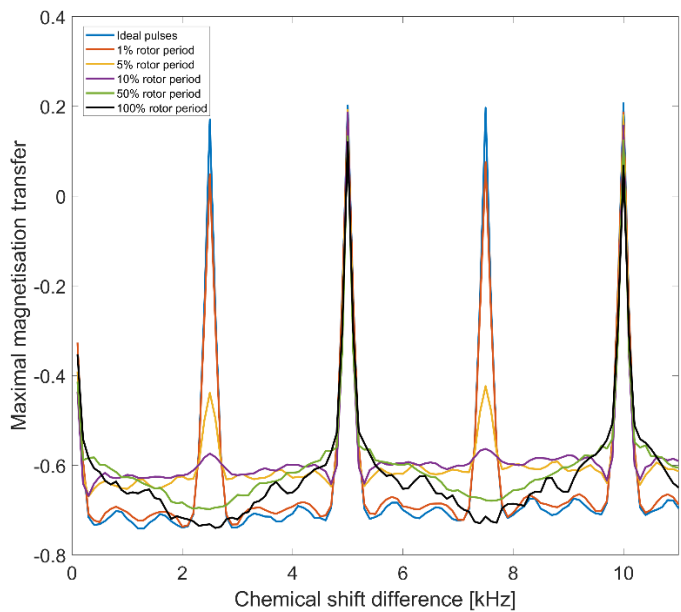

**Figure 3:** Effect of finite 180 pulse length on normalised maximal magnetisation transfer as a function of Δδ at a MAS rate of 5 kHz. The blue curve represents ideal 180° pulses. Pulses with a length of 1%, 5%, 10%, 50% and 100% of the rotor period are represented by orange, yellow, purple, green, and black curves respectively.



**Magnetisation transfer with variable flip angle ¹H pulses:** A unique observation was the dependence of the $\mathcal{HR}2$ condition and PDSD efficiency on the flip angle of the pulses applied on the protons during the mixing time. Figure 4a shows the effects of flip angles between 90° and 180° on the magnetisation transfer at a spinning frequency of 6 kHz. In subplots 4b-d, the maximal magnetisation transfer is presented as a function of both the spinning frequency $\nu_R$ and the chemical shift difference $\Delta\delta$. The results show a trade-off between efficiency of the $\mathcal{HR}2$ condition and a better performance of the enhancement with respect to $\Delta\delta$. Generally, flip angles between 90°-120° exhibited the best total enhancement. However, these angles attenuate the new $\mathcal{HR}2$ condition. The heat-maps (c) and (d) demonstrate that up to ~10 kHz, an enhancement in the polarisation transfer via PDSD is gained by the application of pulses, with a clear advantage to the application of 90° pulses. This enhancement is accompanied by the disappearance of the new $\mathcal{HR}2$ condition.

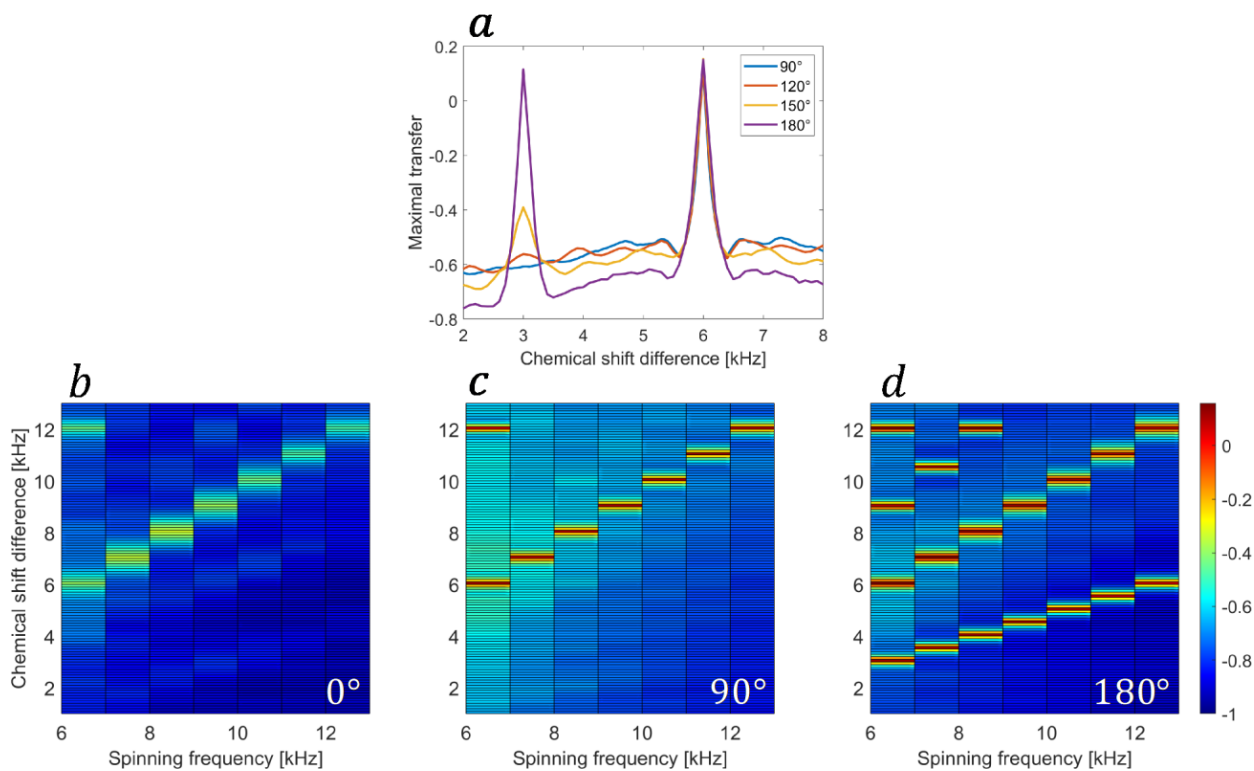

**Figure 4:** (a) Comparison of the normalised maximal magnetisation transfer ($I_{2z}$) as a function of the chemical shift difference $\Delta\delta(^{13}C)$ for different flip angles at $\nu_R$ = 6 kHz. The maximal transfer was taken from the corresponding build-up curves. (b)-(d) are heat-maps showing the maximal magnetisation transfer as a function of $\Delta\delta(^{13}C)$ and $\nu_R$ in the absence of ¹H pulses (b), and in the presence of pulses with flip angles of 90° (c) and 180° (d). The colourmap is normalised such that $I_{2z}(0) = -1$ (dark blue).



Since the new resonance condition and the enhancement originate from the pulses, we term this new scheme 'Pulse Induced Resonance with Angular-dependent Total Enhancement' (PIRATE).

## NMR experiments

In order to experimentally demonstrate the effects shown in the simulations, a $^{13}$C-labelled glycine sample was used as a model system having only two distinctive $^{13}$C chemical shifts. In our 14.1T magnet, the chemical shift difference between the carbonyl ($I_{1z}$) and the alpha carbon ($I_{2z}$) is 20.054 kHz. Therefore, following Eq. 2, the recoupling at the $\mathcal{HR}2$ condition is given by ($n$ odd integer)

$$(3)\ \nu_R = \frac{40.108}{n} [kHz].$$

The $\mathcal{R}2$ condition in this case is fulfilled for even values of $n$. At first, a set of 1D experiments were conducted without any $^1$H pulses, as shown in figure 5a, in order to demonstrate the $\mathcal{R}2$ condition as shown before [19]. A selective inversion pulse was applied to Cα, and $^{13}$C spectra were collected at a longitudinal evolution time $\tau_m$=5 ms. A typical spectrum appears in figure 6a. The magnetisation transfer was quantified by subtracting the integral of the carbonyl signal from that of Cα, giving a value that reports on the expectation value <$I_{1z}$-$I_{2z}$>, as followed in the simulations. This value was then plotted as a function of the MAS rate, as shown in figure 6b (and simulated in figure 2). Then, similar experiments and data analysis were performed by applying a single $^1$H pulse every rotor period as shown in figure 5b.

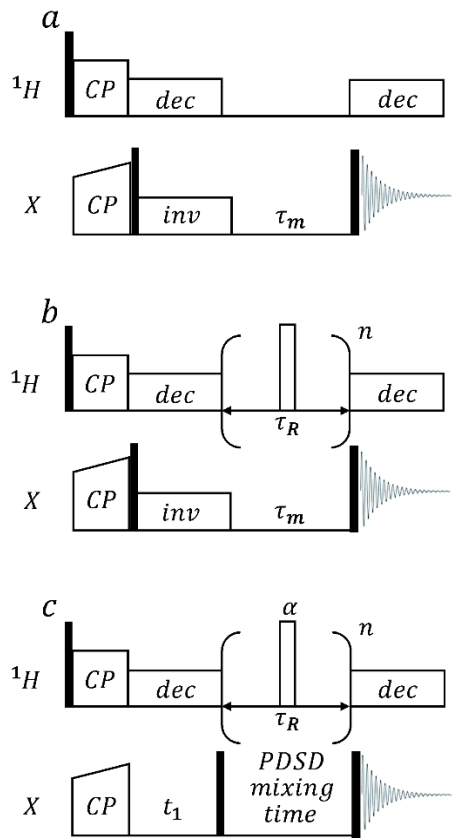

**Figure 5:** Pulse sequences used in our experiments. (a) $^1$H-$^{13}$C Cross-polarization experiment followed by a selective inversion pulse (on Cα), a mixing time, and a detection pulse. (b) As in (a), with rotor-synchronous $^1$H pulses having a flip angle α applied during the mixing time (PIRATE scheme with XY-16 phase cycle). (c) A 2D PIRATE experiment. '*n*' indicates the number of rotor periods.

As expected, in the absence of $^1$H pulses a significant rise in the magnetisation transfer was observed at MAS rates of 10 kHz and 6.5 kHz (n=4,6 in Eq. 3). Additionally, the transfer was more efficient at lower MAS rates, in accordance with decreased averaging of the heteronuclear dipolar interactions. When rotor-synchronous pulses were applied (according to the scheme in figure 5b) at three different flip angles, 90°, 109°, and 180°, transfer efficiencies varied and the new resonance emerged. The results show that the application of 180° pulses on the proton channel during the evolution time created two additional maxima of transfer at 13.369 kHz and at 8 kHz (n=3,5 in equation 3, $\mathcal{HR}2$ condition) that were absent without the application of pulses. When 90° pulses were applied only the $\mathcal{R}2$ condition is met, the enhancement in the PDSD mechanism is visible off-$\mathcal{R}2$ condition, and in particular at the higher regime of MAS rates above 11 kHz, where the efficiency of PDSD in the absence of pulses is significantly smaller. Applying 109° pulses shows enhancement in PDSD compared to the 180° pulses, and the $\mathcal{HR}2$ condition is still visible although at reduced intensity. This result agrees with the simulations above showing the dependence of the resonance on the pulse flip angle.

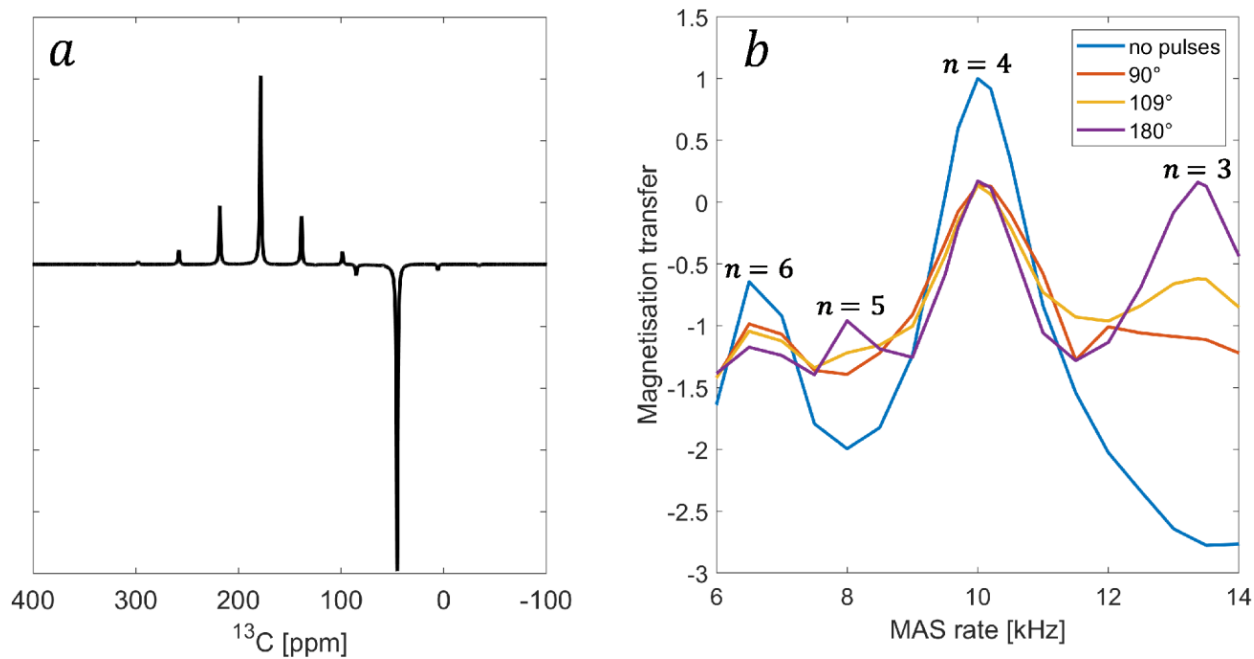

**Figure 6:** (a) A typical spectrum of $^{13}C_2$-glycine taken after inverting C$\alpha$ and allowing magnetisation to transfer for 5 ms using the PIRATE pulse scheme with a flip angle of 180° and a MAS rate of 6 kHz. The spectrum was processed with a line broadening of 200 Hz. (b) Magnetisation transfer as a function of the MAS rate. The transfer was calculated by subtracting the intensity of the carbonyl carbon from that of C$\alpha$ for all spectra acquired as in (a) but using different flip angles or without pulses. The magnetisation transfer was normalised so that the highest signal is 1. The corresponding 'n' value of each MAS rate according to Eq. (3) is indicated in black.

The PIRATE scheme can be applied to multi-dimensional experiments to allow magnetisation transfer. In systems containing spins with dispersed chemical shift differences the $\mathcal{HR}2$ condition can be used to selectively increase the magnetisation transfer. This approach is similar to prior studies utilising the spinning frequency dependence of the $\mathcal{R}2$ condition to perform rotational-resonance width ($\mathcal{R}2_W$) based distance measurements [42] or constant-time frequency-dependent narrow-band radio frequency driven recoupling [21]. In such cases, 180° pulses are highly useful since they enhance the $\mathcal{HR}2$ condition. Moreover, in two-dimensional experiments those selective cross-peaks do not overlap with spinning sidebands. However, for obtaining broad-band enhancement of magnetisation transfer that is independent of $\Delta\delta$, 90°-120° pulses are superior.

A set of 2D PIRATE experiments, as depicted in figure 5c, was conducted to demonstrate the enhancement in multidimensional experiments. The experiments were conducted on $^{13}C_2$- glycine at the $\mathcal{HR}2$ condition ($\nu_R$=13.369 kHz) and again at off resonance conditions ($\nu_R$=11.5 kHz) with different flip angles. The signal to noise ratios (SNR) of the diagonal and correlation signals were used for comparison and appear in Table 1. The efficiency of transfer was determined by the ratio between the SNR of the correlation cross-peak to the sum of both diagonal and correlation cross-peaks.

Figure 7a shows a comparison of on- and off-$\mathcal{HR}2$ resonance 2D experiments using 180° pulses. A clear enhancement in the intensity of the cross-peaks at half-rotational-resonance conditions is depicted in the plot. The enhancement in transfer efficiency off-resonance due to the different flip angles in the PIRATE scheme is indicated by an increase in the relative SNR when 90° pulses are applied instead of 180° pulses, as demonstrated in figure 7b. On-resonance, as shown in figure 7c and quantified in table 1, the spectra from 90° pulses exhibit a decrease in the SNR values with respect to 180°, demonstrating the dependence of the $\mathcal{HR}2$ condition efficiency on the flip angle. Moreover, this also shows the applicability of this experiment for moderate MAS rates where the PDSD experiment is less efficient. Overall, the values are in agreement with simulations although with reduced efficiencies, as will further be discussed below. The transfer efficiencies when the $\mathcal{HR}2$ condition is met are superior to all other conditions, are higher than the total enhancement of the 90° pulses, and the latter are superior at off-resonance conditions.

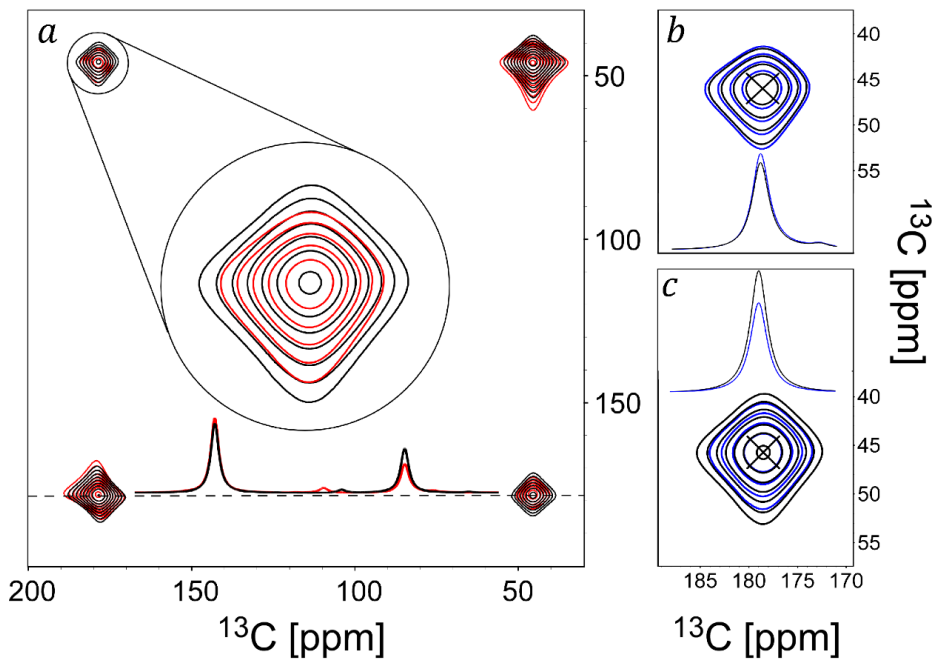

**Figure 7:** Comparison of 2D $^{13}$C-$^{13}$C correlation experiments using the PIRATE scheme. (a) $^1$H pulses of 180° pulses were applied on-resonance (13,369 Hz, black) and off-resonance (11,500 Hz, red). The enhancement in the intensity due to the $\mathscr{HR}2$ condition is demonstrated by the enlarged Cα-C correlation cross-peak and by the 1D slice taken at the position of the dashed line. (b) Cα-C cross-peak intensity at off-resonance conditions with 180° (black) and 90° pulses (blue), along with its corresponding 1D slice. (c) Cα-C cross-peak at on-resonance with 180° (black) and 90° pulses (blue), along with its 1D slice. All spectra were processed with line broadening of 500Hz on both dimensions and are drawn using an identical contour level scale. The lowest contour is drawn at 5 times the SNR and subsuqeunt contours are at multiples of 1.40.

| Cross-peak | Flip angle | Off-resonance (11,500Hz) | On-resonance (13,369Hz) | Transfer efficiency |
|---|---|---|---|---|
| Cα-Cα | 90° | 1.00 | 0.87 | |
| | 180° | 0.80 | 0.67 | |
| C-C | 90° | 0.78 | 0.77 | |
| | 180° | 0.63 | 0.59 | |
| Cα-C | 90° | 0.35 | 0.30 | 25.9% |
| | 180° | 0.24 | 0.36 | 35.2% |
| C-Cα | 90° | 0.35 | 0.31 | 30.7% |
| | 180° | 0.24 | 0.38 | 38.9% |

**Table 1:** SNR values of cross-peaks from PIRATE 2D spectra collected with flip angles of 90° and 180° at MAS rates on- and off- half-rotational-resonance. The SNR values were normalised with respect to the Cα-Cα diagonal peak derived from the experiment with α=90°. Transfer efficiency refers to the ratio between the correlation signal to the sum of that correlation cross-peak and the matching diagonal signal. For 180° pulses it is calculated for the on-resonance MAS rate (indicated by blue cells), whereas for the 90° pulses it is calculated for the off-resonance MAS rate (indicated by green cells).

In order to further demonstrate the selectivity of the experiment on a larger spin system, and compare with current techniques, we performed 2D PDSD, PIRATE 180° and DARR experiments on a U-$^{13}$C/$^{15}$N labeled fd-Y21M filamentous bacteriophage virus. Figure 8 shows 1D slices extracted at the Cα chemical shift of valine 33 (V33) [38] . The intensity of the diagonal peak, corresponding to the $C_\alpha$ chemical shift, was normalised to one in the three experiments. The MAS rate was chosen to be $v_R = 10 kHz$, so that the chemical shift difference between the V33$C_\alpha$ (66.0 ppm) and V33$C_\beta$ (31.6 ppm) matches the n=1 condition in equation 2. It is visible that both PIRATE and DARR show an enhancement compared to the PDSD experiment. Moreover, the transfer of signal from the diagonal peak to the other spins is more significant in DARR than in PIRATE, with the exception of the $C_\alpha - C_\beta$ transfer, meaning that the PIRATE experiment selectively enhanced the magnetisation transfer between those two spins. This selectivity could also be spotted with alanine 35 $C_\alpha - C_\beta$, that shares a similar chemical shift difference to V33, and is shown in figure S4 in the SI. Even more pronounced, is the difference in transfer efficiency to the carbonyl carbons located away from the resonance condition ($\delta_{V33C'}$=178.3 ppm, i.e. $\Delta\delta_{V33C'-C\alpha}$=16.9 kHz whereas at n=3 the condition is at 15 kHz).

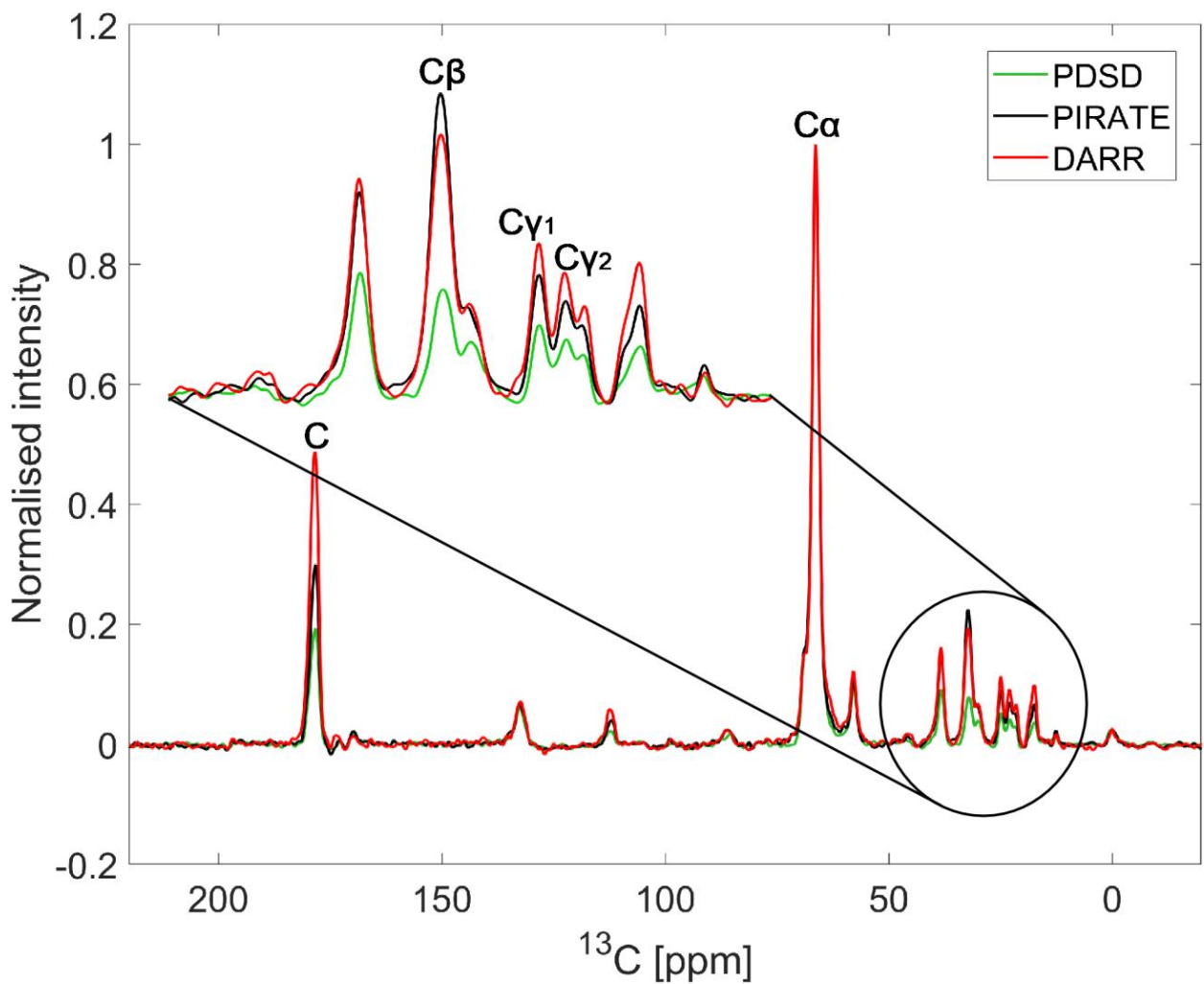

**Figure 8:** Comparison of 1D slices (F1 = 66.0 ppm), taken from 2D experiments on fd-Y21M bacteriophage. The green, black and red lines represent PDSD, PIRATE 180° and DARR experiments respectively. The assignment for valine 33 is denoted next to the relevant signals. The spinning rate for all three experiments was 10 kHz. The spectrum was processed with Lorentz to Gauss (inverse exponential -50Hz, Gaussian maximum position 0.1) on both dimensions. The lowest contour is drawn at 5 times the SNR and subsuqeunt contours are at multiples of 1.40.

## Summary and Conclusions

We have demonstrated the existence of a "half rotational resonance" condition in solid-state NMR experiments involving homonuclear spins coupled to protons. When the MAS rate is fixed to half (and full) integers of the chemical shift difference between the two coupled spins, a significant enhancement in the transfer occurs if rotor-synchronous pulses are applied to the proton channel. In addition to the enhancement at this new resonance condition, an improvement in polarisation transfer was observed across all chemical shift difference values (broad-band enhancement) in comparison to PDSD yet smaller than DARR. The flip angle of the pulses was shown to have an impact on both effects, where for an angle of 180° the new resonance condition is maximised and for angles closer to 90° the PDSD effect is superior across a broad range of chemical shift difference values.

From simulations we could reveal the key terms in the spin Hamiltonian that contribute to the effect. Half-rotational resonance was observed for a minimal H-X-X system containing a single proton coupled to some other spin X via heteronuclear dipolar coupling, whereas the X spin is homonuclear-coupled to another X spin. The $^1$H-$^1$H homonuclear interactions are partially recoupled by the application of the pulses providing broad-band increase of the PDSD effect. We find that X-spin CSA has a very small effect on transfer efficiency at values relevant to $^{13}$C and $^{15}$N spins.

Through experiments on $^{13}$C$_2$-glycine we demonstrated the enhancement at the resonance condition when using PIRATE with 180° pulses, and broad-band enhancement when using 90° pulses. Experiments on an intact virus showed the selectivity of PIRATE 180°, which can be utilised for selective transfer in multidimensional experiments in more complex systems. In both systems, the intensity at the resonance condition was not as pronounced as in simulations. This observation is probably due to several differences; (i) in simulations (and in 1D experiments) the initial condition was an inverted spin unlike 2D experiments, (ii) a minimal spin system was used for simulations while in the experiments we have a fully $^{13}$C labelled system, and therefore magnetisation can transfer to more remote spins that are not accounted for, (iii) we report the maximal transfer in simulations while experiments have been performed at a fixed mixing time, and (iv) experiments use finite pulses that reduce the resonance efficiency as discussed above.

This work, and specifically the resonance condition, can be further generalised. If we assume that a 180° $^1$H pulse inverts some RF toggling frame Hamiltonian, and the next pulse inverts it back, the heteronuclear interaction is modulated by a period of 2T$_R$, or ν$_R$/2. Thus, a spacing of an integer 'k' rotor periods between pulses will result in a different modulation of the heteronuclear interactions, producing more resonances of the form $\Delta\delta = \frac{n}{2k}\nu_R$. The efficiency of such an approach is yet to be explored and will be described in more details in future work.

## Acknowledgments

This research was support by the Israel Science Foundation grant #847/17. We would like to thank Dr. Gili Abramov for the preparation of the fd-y21m phage sample. We would also like to thank reviewer #2 for insightful and enriching comments.

# Supporting Information

## Pulse Induced Resonance with Angular Dependent Total Enhancement of multi-dimensional solid-state NMR correlation spectra

Orr Simon Lusky[1], Amir Goldbourt[1*]

[1]*School of Chemistry, Faculty of Exact sciences, Tel Aviv University, Tel Aviv, Israel.*

*amirgo@tauex.tau.ac.il

**Simpson script for PIRATE:** in red – parameters that were varied in simulations

```
spinsys {
    channels 13C 1H
    nuclei 13C 13C 1H 1H 1H
    shift 1 -1000 2p 0.0 -29.6 91.2 -19.4
    shift 2 5000 10p 0.0  29.6 21.2 -19.4
    dipole 1 2 -2000 0 0 0
    dipole 1 3 -21940 0 80 0
    dipole 1 4 -21940 0 70 0
    dipole 1 5 -21940 0 60 0
    dipole 2 3 -21940 0 10 0
    dipole 2 4 -21940 0 30 0
    dipole 2 5 -21940 0 20 0
    dipole 3 4 -25040 0 50 0
    dipole 3 5 -25040 0 150 0
    dipole 4 5 -55040 0 40 0
}
par {
    start_operator     I1z-I2z
    detect_operator    I2z
    spin_rate          [6000-13000]
    gamma_angles       20
    sw                 spin_rate
    crystal_file       rep100
    np                 128
    verbose            1101
    proton_frequency   400e6
    conjugate_fid true
}
proc pulseq {} {
   global par
   set tr [expr 1.0e6/$par(spin_rate)]
   set tr2 [expr 0.5e6/$par(spin_rate)]

delay $tr2
delay $tr2
```

```
pulseid 5 0000 0 100000 0
store 1
reset
   for {set i 0} {$i<$par(np)    } {incr i} {
   acq
   prop 1
   }
}
proc main {} {
    global par
    set f [fsimpson]
    fsave $f $par(name).fid -xreim
}
```

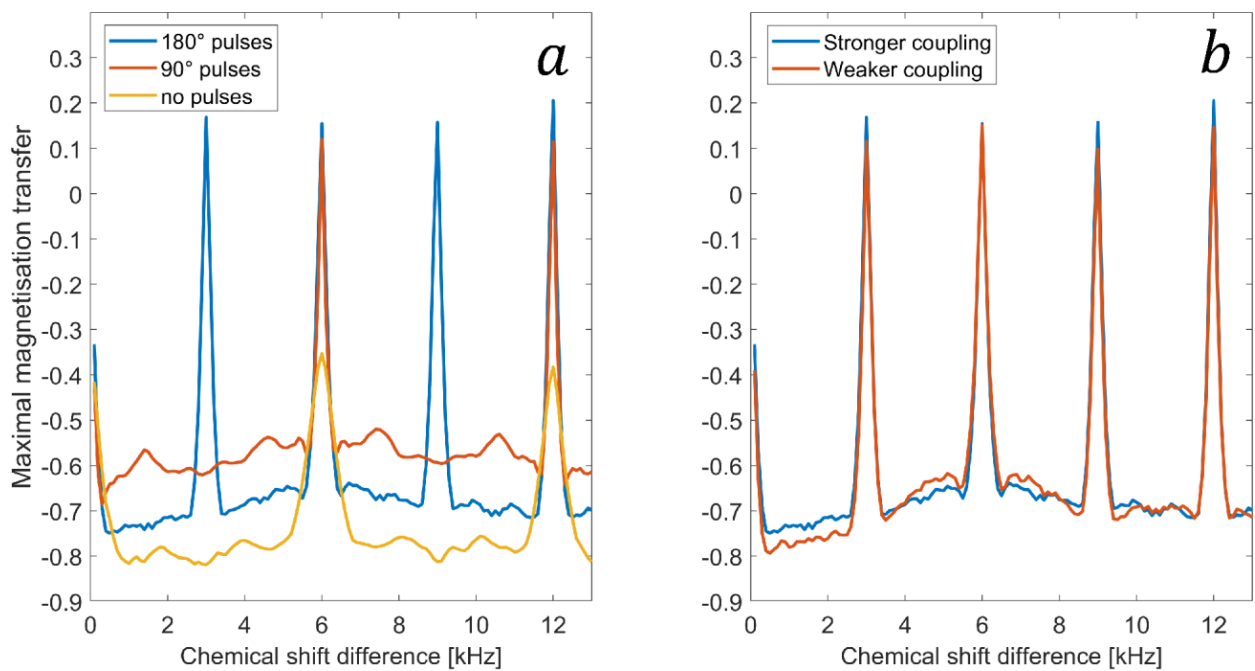

**Figure S1**. Simulations of a fully coupled $C_2H_3$ spin system with reduced coupling ($D_{H4-H5}$=30 kHz; 1.6Å as in $CH_2$ group) between the protons 4 and 5 in the script above. (a) Application of the PIRATE scheme with 180°, 90° and no pulses using $D_{H4-H5}$=30 kHz. (b) Comparison of magnetisation transfer in PIRATE with stronger coupling ($D_{H4-H5}$=30 kHz) and weaker coupling ($D_{H4-H5}$=55.04 kHz).

**Figure S2.** Build-up curves at $\nu_R = 6\ kHz$ showing the normalised intensity of an inverted spin as a function of mixing time at rotational-resonance (blue, '$\mathcal{R}2$', $\Delta\delta = n\nu_R = 6kHz$), half-rotational resonance (orange, '$\mathcal{HR}2$', $\Delta\delta = 0.5n\nu_R = 3kHz$), and off resonance (yellow, 'off', $\Delta\delta = 5kHz$). Connectivities between the $^1$H spins (green circles) and $^{13}$C spins (blue circles) represent the dipolar coupling networks. Plot '0' is the fully coupled system simulated at the three conditions above and represents the blue curves in all subplots of figure 2. Similarly, subplots S2a-h each consists of three examples of the build-up curves that were used to generate subplots a-h in figure 2. The couplings were as follows: (0, a-e) $D_{^{13}C-^{13}C} = 2$ kHz, $D_{^1H-^{13}C} = 21.94$ kHz, $D_{^1H-^1H} = 25.04$ kHz twice and 55.04 kHz once; (f) $D_{^{13}C-^{13}C} = 1$ kHz; (g) the $\beta$ angles of the dipolar tensors were varied (pairs of different X-$^1$H spins both set to 60°, two pairs of $^1$H-$^1$H, both $D_{^1H-^1H} = 25.04$ kHz, set to 40°); (h) CSA of $^{13}$C are 200 ppm/300 ppm; (i) $D_{^{13}C-^{13}C} = 0$kHz;

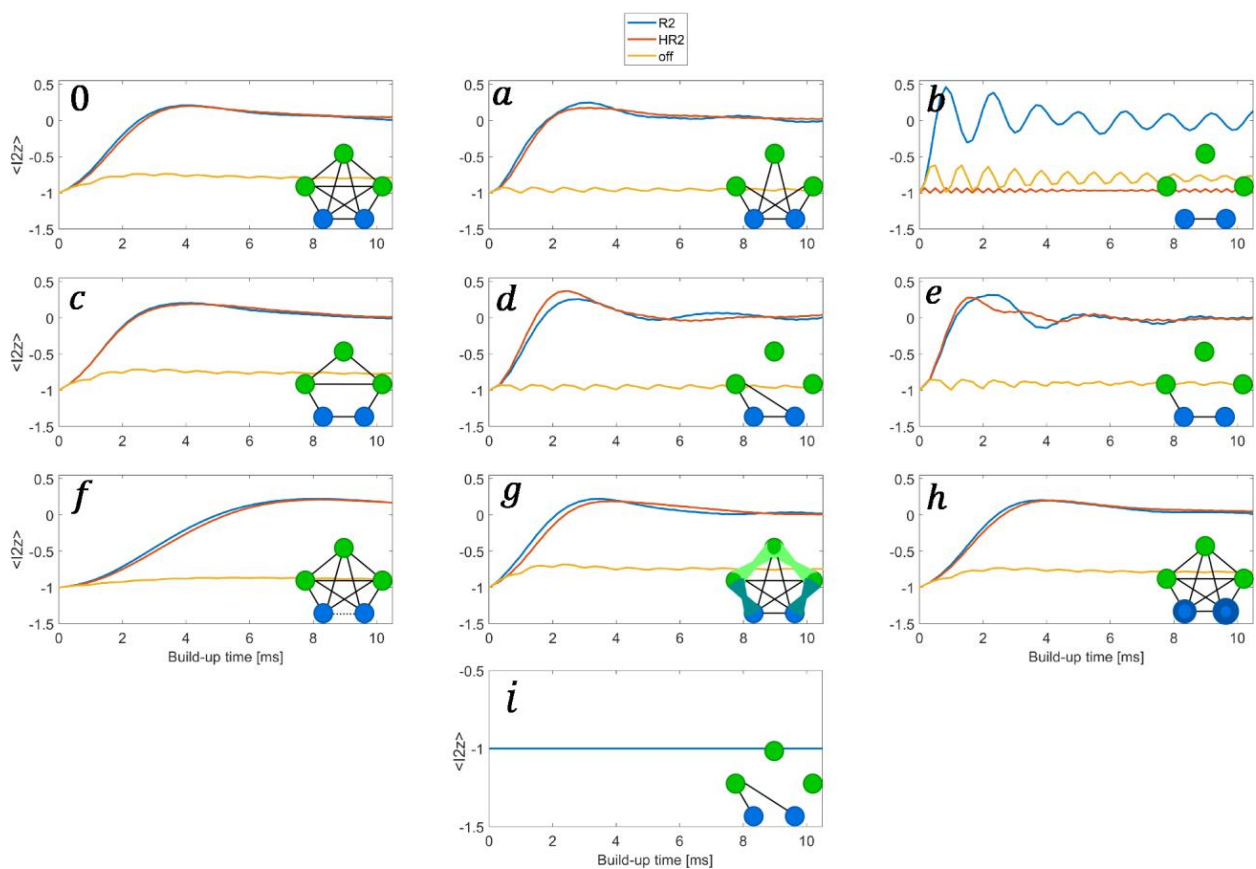

**Figure S3.** Simulations of $I_{1z}$ (blue), $I_{2z}$ (orange), and their sum $I_{1z}+I_{2z}$ (yellow) for three different cases at $\nu_R = 6\ kHz$. (a) Application of 180° pulses at the $\mathcal{HR}2$ condition ($\Delta\delta = 3kHz$) on the full system; (b) Application of 90° pulses off-resonance ($\Delta\delta = 4kHz$) on the full $C_2H_3$ system; (c) Application of 180° pulses on the system described in S2e (H-C-C) off resonance ($\Delta\delta = 4kHz$).

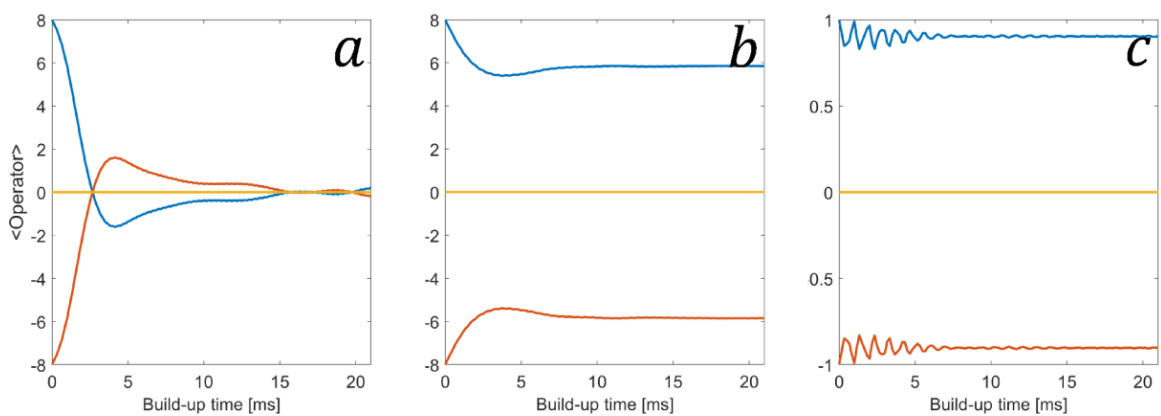

**Figure S4.** Comparison of 1D slices (F1 = 55.1 ppm), taken from 2D experiments on fd-Y21M bacteriophage at $\nu_R = 10\ kHz$. The green, black and red lines represent PDSD, PIRATE 180° and DARR experiments respectively. The assignment for alanine 35 is denoted next to the relevant signals and are magnified. The chemical shift difference between $A35C_\alpha$ (55.1ppm) and $A35C_\beta$ (21.4ppm) is 5.08kHz, a value that matches the n=1 condition in $\Delta\delta = \frac{n}{2}\nu_R$. The chemical shift difference between $A35C_\alpha$ and $A35C'$ (179.0ppm) is 18.68kHz and is closest to the condition n=4, which is 20kHz. The spectrum was processed with Lorentz to Gauss (inverse exponential -50Hz, Gaussian maximum position 0.1) on both dimensions.

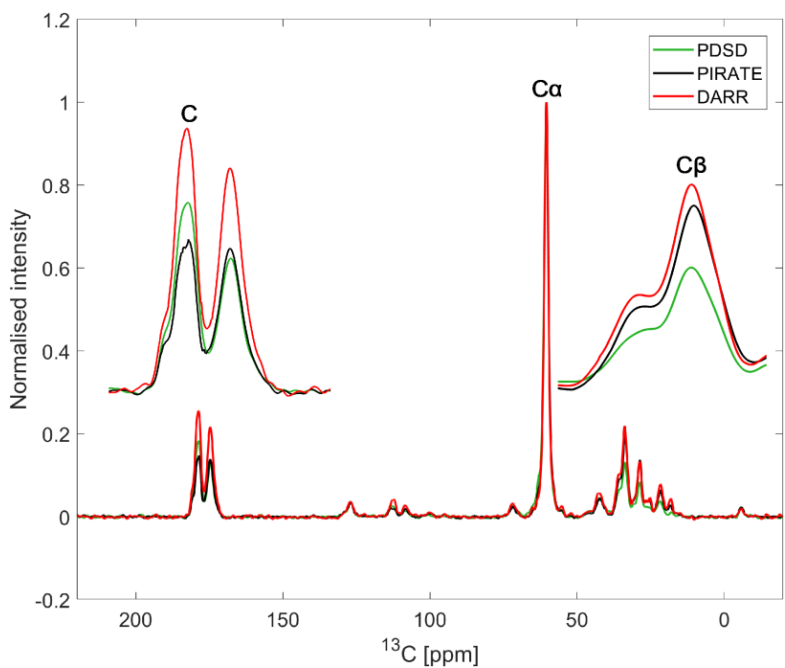

**Experimental parameters:**

**Table S1a:** Parameters for 1D $^{13}$C-$^{13}$C CP-selective inverse experiments

| Experiment group | PDSD | PIRATE 90° | PIRATE 109.5° | PIRATE 180° |
|---|---|---|---|---|
| Field [T] | 14.1 | | | |
| Spinning frequency ($\nu_R$) [kHz] | 5, 5.5, 6, 6.5, 7, 7.5, 8, 8.5, 9, 9.5, 9.7, 10, 10.2, 10.5, 11, 11.5, 12, 12.5, 13, 13.4, 13.5, 14, 14.5, 15 | | | |
| Acquisition points | 3988 | | | |
| Acquisition time [ms] | 20 | | | |
| Carrier frequency [ppm] | 111 | | | |
| Frequency jump (for selective inverse pulse) [Hz] | -9680 | | | |
| $^1$H 90° pulse length [µs] | 2.5 | | | |
| $^{13}$C 90° pulse length [µs] | 4 | | | |
| $^1$H PIRATE pulses length [µs] | No pulses | 2.5 | 3.04 | 5 |
| $^{13}$C 180° inversion pulse length [µs] | 100 | | | |
| CP power ($\nu_H$) [kHz] | 77.5 | | | |
| CP power ($\nu_C$) [kHz] | 62.5 | | | |
| CP contact time [ms] | 2 | | | |
| PIRATE mixing time [ms] | 5 | | | |
| $^1$H decoupling power [kHz] | 100 | | | |
| swf-tppm decoupling pulse [µs] | 5 | | | |
| Relaxation delay [s] | 2.4 | | | |
| Scans | 16 | | | |
| SW [kHz] | 100 | | | |
| **Processing parameters** | | | | |
| Processing software | TopSpin 3.5 | | | |
| Zero fill | 8192 | | | |
| Apodization | Line broadening 200Hz | | | |

**Table S1b:** Parameters for 2D $^{13}$C-$^{13}$C PIRATE experiments conducted on $^{13}$C$_2$-glycine

| Experiment | Off-resonance PIRATE 90° | Off-resonance PIRATE 180° | On-resonance PIRATE 90° | On-resonance PIRATE 180° |
|---|---|---|---|---|
| Field [T] | 14.1 | | | |
| Spinning frequency ($\nu_R$) [kHz] | 11.5 | 11.5 | 13.369 | 13.369 |
| Acquisition points (t1/t2) | 3988/450 | | | |
| Acquisition time [ms] (t1/t2) | 0.02/0.0045 | | | |
| Carrier frequency [ppm] | 111 | | | |
| $^1$H 90° Pulse length [µs] | 2.5 | | | |
| $^{13}$C 90° Pulse length [µs] | 4.5 | | | |
| PIRATE pulses length [µs] | 2.5 | 5 | 2.5 | 5 |
| CP power ($\nu_H$) [kHz] | 74 | | | |
| CP power ($\nu_C$) [kHz] | 55.5 | | | |
| CP contact time [ms] | 2 | | | |
| PIRATE mixing time [ms] | 5 | | | |
| $^1$H decoupling power [kHz] | 100 | | | |

| | |
|---|---|
| swf-tppm decoupling pulse [μs] | 5 |
| Relaxation delay [s] | 2.4 |
| Scans | 16 |
| SW F1/F2 [kHz] | 100/50 |
| **Processing parameters F1/F2** | |
| Processing software | TopSpin 3.5 |
| Zero fill (t1/t2) | 8192/1024 |
| Apodization | Line broadening 500Hz on both dimensions |

**Table S1c:** Parameters for 2D $^{13}$C-$^{13}$C PIRATE and DARR experiments conducted on fully labelled $^{13}$C, $^{15}$N fd-Y21M bacteriophage.

| Experiment | PIRATE 180° | DARR |
|---|---|---|
| Field [T] | 14.1 | |
| Spinning frequency ($\nu_R$) [kHz] | 10 | |
| Acquisition points (t1/t2) | 3000/512 | |
| Acquisition time [ms] (t1/t2) | 15/10 | |
| Carrier frequency [ppm] | 85 | |
| $^1$H 90° Pulse length [μs] | 2.8 | |
| $^{13}$C 90° Pulse length [μs] | 4.5 | |
| PIRATE pulses length [μs] | 5.6 | - |
| CP power ($\nu_H$) [kHz] | 75.5 | |
| CP power ($\nu_C$) [kHz] | 55.5 | |
| CP contact time [ms] | 1 | |
| $^1$H power during DARR [kHz] | - | 10 |
| Mixing time [ms] | 10 | |
| $^1$H decoupling power [kHz] | 80 | |
| swf-tppm decoupling pulse [μs] | 9 | |
| Relaxation delay [s] | 2.6 | |
| Scans | 32 | |
| SW F1/F2 [kHz] | 100/25.6 | |
| **Processing parameters F1/F2** | | |
| Processing software | TopSpin 3.5 | |
| Zero fill (t1/t2) | 8192/1024 | |
| Apodization | Lorentz to Gauss on both dimensions (inverse exponential 50Hz, Gaussian maximum position 0.1) | |